# Large perpendicular magnetic anisotropy in Ta/CoFeB/MgO on full coverage monolayer MoS$_2$ and first principle study of its electronic structure

Ziqi Zhou[1,2], Paul Marcon[3], Xavier Devaux[2], Philippe Pigeat[2], Alexandre Bouché[2], Sylvie Migot[2], Abdallah Jaafar[2], Remi Arras[3], Michel Vergnat[2], Lei Ren[4], Hans Tornatzky[4], Cedric Robert[4], Xavier Marie[4], Jean-Marie George[5], Henri-Yves Jaffrès[5], Mathieu Stoffel[2], Hervé Rinnert[2], Zhongming Wei[1#], Pierre Renucci[4‡], Lionel Calmels[3†], Yuan Lu[2*]

[1]*State Key Laboratory of Superlattices and Microstructures, Institute of Semiconductors, Chinese Academy of Sciences & Center of Materials Science and Optoelectronics Engineering, University of Chinese Academy of Sciences, Beijing 100083, China*
[2]*Institut Jean Lamour, Université de Lorraine, CNRS UMR7198, Campus ARTEM, 2 Allée André Guinier, BP 50840, 54011 Nancy, France*
[3]*CEMES, CNRS, Université de Toulouse, 29 rue Jeanne Marvig, BP 94347, F-31055 Toulouse, France*
[4]*Université de Toulouse, INSA-CNRS-UPS, LPCNO, 135 Av. Rangueil, 31077 Toulouse, France*
[5]*Unité Mixte de Physique, CNRS, Thales, Université Paris-Saclay, 91767 Palaiseau, France*

Corresponding authors: *yuan.lu@univ-lorraine.fr; †lionel.calmels@cemes.fr; ‡renucci@insa-toulouse.fr; #zmwei@semi.ac.cn

**Abstract**

Perpendicularly magnetized spin injector with high Curie temperature is a prerequisite for developing spin optoelectronic devices on 2D materials working at room temperature (RT) with zero applied magnetic field. Here, we report the growth of Ta/CoFeB/MgO structures with a large perpendicular magnetic anisotropy (PMA) on full coverage monolayer (ML) MoS$_2$. A large perpendicular interface anisotropy energy of 0.975mJ/m$^2$ has been obtained at the CoFeB/MgO interface, comparable to that observed in magnetic tunnel junction systems. It is found that the insertion of MgO between the ferromagnetic metal (FM) and the 2D material can effectively prevent the diffusion of the FM atoms into the 2D material. Moreover, the MoS$_2$ ML favors a MgO(001) texture and plays a critical role to establish the large PMA. First principle calculations on a similar Fe/MgO/MoS$_2$ structure reveal that the MgO thickness can modify the MoS$_2$ band structure, from an indirect bandgap with 7ML-MgO to a direct bandgap with 3ML-MgO. Proximity effect induced by Fe results in a splitting of 10meV in the valence band at the Γ point for the 3ML-MgO structure while it is negligible for the 7ML-MgO structure. These results pave the way to develop RT spin optoelectronic devices on 2D transition-metal dichalcogenide materials.

**INTRODUCTION**

Transition metal dichalcogenides (TMDs) have emerged as a promising 2D crystal family, which may open the routes for novel nano-electronic and opto-electronic device applications[1,2,3,4,5,6,7]. In contrast to graphene and boron nitride (BN), which are respectively a semi-metal and a wide-gap semiconductor, the family of TMDs displays a large variety of electronic properties ranging from semi-conductivity to superconductivity[8]. As a representative of TMDs, molybdenum disulfide ($MoS_2$) has a tunable bandgap that changes from an indirect gap of 1.2 eV in the bulk crystal to a direct gap of 1.8 eV for one monolayer (ML)[1]. The $MoS_2$ ML is characterized by a large spin-orbit splitting of ~0.15 eV for the valence band[3,4], together with a small value of ~3 meV for the conduction band[9]. The lack of inversion symmetry combined with the spin-orbit interaction leads to a unique coupling of the spin and valley degrees of freedom, yielding to a robust spin and valley polarization[4,5,6,7].

To obtain a spin optoelectronic device[10,11,12,13] based on a $MoS_2$ monolayer with zero applied magnetic field, one of the prerequisites is to realize a robust spin injector or detector with perpendicular magnetic anisotropy. Two main reasons explain this prerequisite. First, for a $MoS_2$ ML, the intrinsic spin splitting of both the conduction and valence bands due to the interplay between the breaking of the inversion symmetry along the growth direction and the spin-orbit coupling (SOC) favor the spin transport through $MoS_2$ with an out-of-plane spin polarization[14]. This is because the SOC creates an effective $k$-dependent magnetic field perpendicular to the layer due to the Dresselhaus interactions[15,16]. If electrons with in-plane spin polarization are injected into a $MoS_2$ ML, the perpendicular effective magnetic field can induce an efficient in-plane spin precession along the field due to the D'yakonov-Perel (DP) spin relaxation mechanism[17,18] as well as a spin-dephasing. Consequently, this yields a predicted short spin lifetime (10-200 ps)[19] together with a small spin diffusion length (~20 nm)[20]. Second, according to the optical selection rules[21] in 2D materials, the magnetization of the spin injector has to be maintained perpendicular to the sample surface in order to emit a circularly polarized light from the surface emission geometry. This usually requires a strong external magnetic field up to several Tesla to keep the magnetization perpendicular, which is not favorable for practical applications. Hole spin injection into a TMDs ML has been recently demonstrated by electrical injection and optical detection method, using either a perpendicularly magnetized GaMnAs injector at zero applied magnetic field[11] or an in-plane magnetized NiFe injector at large perpendicular magnetic field[10].

Different ways can be used to get a ferromagnetic (FM) layer exhibiting a PMA on 2D materials. One way is to build FM 2D/TMD Van der Waals (VdW) heterojunctions by using 2D transfer technology with FM 2D materials showing PMA such as $CrI_3$[22], $Fe_3GeTe_2$[23] and Fe-doped $SnS_2$[24]. This method allows to obtain a sharp interface and avoids interdiffusion between the different layers. However, the Curie temperature ($T_C$) of these materials is often below room temperature and the size of the junctions is too small (around tens of microns) due to the exfoliation procedure. Another way is to use a high Curie temperature $3d$ FM metal, such as CoFeB[25] or Co[26], on top of a 2D material. However, the direct deposition of FM metals on 2D materials often leads to metal diffusion into the 2D materials, and also changes the electronic structure of the 2D materials from semiconducting to metallic properties[27,28,29], due to covalent bonds between atoms of the metals and of the 2D materials that destroy the bandgap[29]. A good alternative way is to insert a thin insulating layer between the FM and the 2D material. This can on one hand avoid the interdiffusion problem and on the other hand preserve the bandgap of 2D materials. Furthermore, the insertion of a thin insulating layer can tune the interface resistance and circumvent the conductivity mismatch between the metal and the semiconductor for efficient spin injection[30,31].

It is well known that Ta/CoFeB/MgO exhibits a strong perpendicular magnetic anisotropy, due to its interfacial anisotropy energy. The origin of this PMA is due to the hybridization of Co(Fe) and O atomic orbitals at the CoFeB/MgO interface[32]. This layer stack has been already used in metallic magnetic tunnel junctions (MTJs)[33] and on spin light emitting diode (spin LED)[34,35,36,37,38] based on III-V material (GaAs). However, the growth of the Ta/CoFeB/MgO structure on 2D materials has never been reported until now. In this work, we report on the elaboration of a Ta/CoFeB/MgO/$MoS_2$ heterostructure with a strong PMA on full coverage $MoS_2$ ML. Compared with the standard procedure of the 2D exfoliation, the successful fabrication of this heterostructure will have important impacts. First, it will allow to fabricate high density spintronics devices with UV lithography on a large substrate. Second, it will preserve clean and sharp CoFeB/MgO and MgO/MoS$_2$ interfaces since we can use a top-to-down lithography procedure to fabricate the device. This is especially important for achieving efficient spin injection[31,39] and for resolving the contact problem between the FM and 2D layers[40].

**RESULTS AND DISCUSSION**

*Characterization of the MoS$_2$ monolayer*

**Figure 1a** schematically shows the prepared heterostructure consisting of SiO$_2$ sub.//MoS$_2$ ML /MgO (1.5 nm) /CoFeB(1.2 nm)/Ta (5 nm). The MoS$_2$ ML was grown by chemical vapor deposition (CVD) on SiO$_2$ substrate. The uniformity of the MoS$_2$ was firstly checked by optical microscopy. A uniform contrast is observed on the whole surface (1×1cm$^2$) (**Figure 1b**). To further prove the monolayer character and to verify the homogeneity of the MoS$_2$ layer, we have performed characterizations with photoluminescence (PL) and Raman spectroscopy at room temperature (RT). **Figure 1c** shows the PL spectra measured at different positions on the film surface. A strong single peak centered at 1.84eV is obtained for five different locations thus demonstrating a good homogeneity/uniformity of the MoS$_2$ layer. Moreover, the peak position corresponds to the typical direct bandgap energy expected for a single monolayer of MoS$_2$. **Figure 1d** shows the Raman spectra recorded using a laser excitation wavelength of 325 nm. We can identify several peaks corresponding to different phonon scattering modes. The wavenumber difference between the $E_{2g}^1$ mode and the $A_{1g}$ mode is equal to 20.79 cm$^{-1}$ (inset of **Figure 1d**), as expected for the character of MoS$_2$ ML[41]. Reflection high energy electron diffraction (RHEED) has been performed to verify the crystallinity of the MoS$_2$ ML after annealing at 150°C in a molecular beam epitaxy (MBE) chamber. The RHEED pattern shown in the inset of **Figure 1b** displays a sharp streaky character indicating high crystalline quality and a very smooth surface of the MoS$_2$ ML. Since the RHEED pattern does not change by varying the incident angle of the electron beam, we can thus conclude that the MoS$_2$ monolayer remains textured with some preferential orientations in the plane[42].

*Interface structure and chemical characterization of the heterostructure*

High resolution transmission electron microscopy (HRTEM) and scanning transmission electron microscopy (STEM) were performed to characterize the interfacial structure. **Figure 2a** displays the STEM bright field image of the Ta/CoFeB/MgO/MoS$_2$ heterostructure after a rapid thermal annealing (RTA) at 300°C. From the image contrast, we can distinguish different layers. On the top surface of the sample, a 2 nm thick TaO$_x$ layer is formed due to the oxidation of the Ta layer. The thickness of the Ta and CoFeB layers is about 4.8 nm and 1.2 nm, respectively. The interface between Ta and CoFeB is difficult to precisely distinguish because both layers exhibit amorphous features. On the contrary, some crystalline structures can be found in the MgO layer. Beneath the MgO layer, one can identify the MoS$_2$ ML showing a dark contrast. **Figure 2b** shows a magnified STEM image where the MgO layer exhibits a textured structure. Since MgO has a Rock-Salt *fcc* structure, the observed MgO planes could correspond either to MgO (001) or to MgO (111) planes. To clarify this point, we have extracted the profile of the integrated intensity in the green area defined in **Figure 2b**. As shown in **Figure 2c**, the intensity profile shows regular oscillations. Since the dark contrast in the bright field image corresponds to the atomic planes, the distance between two atomic planes is given by the distance separating two successive oscillation minima. The latter is plotted as a function of the interval number (inset of **Figure 2c**). Interestingly, the distance decreases from about 0.4 nm (first interval) to 0.21 nm (sixth interval). The large distance of 0.4 nm between the first MgO plane and the MoS$_2$ ML could be due to the VdW bonding between MgO and MoS$_2$. When the distance from the MoS$_2$ layer increases, the interval reaches a constant value of 0.21 nm, which is consistent with the distance between MgO (001) planes (0.21 nm) but not with MgO (111) planes (0.243 nm). Therefore, we can thus conclude that the MgO is textured with the formation of (001) planes on the MoS$_2$ ML.

In order to understand the inter-diffusion induced by the annealing process, the interfacial element distribution analyses were performed by STEM combined with spatially resolved electron energy loss spectroscopy (EELS) characterizations. **Figures 2d-k** show semi-quantitative chemical maps drawn by processing the EELS spectrum images. The clear layered structures of S, Fe, Co and O prove that there is no evident interdiffusion of elements from the FM layer into the $MoS_2$ layer, thus demonstrating that MgO acts as an efficient diffusion barrier. As expected, the thickness of the Mo layer is thinner than that of the S layer and sites at the center of $MoS_2$ layer, which again validates that the structure of $MoS_2$ is kept unchanged even after annealing. Moreover, B atoms are found to largely diffuse into Ta, which could be due to the good affinity of Ta for B in this system. This is critical to establish the PMA property at the MgO/CoFeB interface[35,36], as will be discussed below.

*Perpendicular magnetic anisotropy of Ta/CoFeB/MgO on $MoS_2$*

The advantage of the full coverage monolayer $MoS_2$ is that it allows us to characterize the magnetic properties of the sample directly by superconducting quantum interference device (SQUID) magnetometry. To obtain an ultrathin CoFeB layer with PMA on $MoS_2$, we have optimized the CoFeB thickness and the annealing temperature ($T_{an}$). **Figure 3a** displays the out-of-plane magnetization *vs.* external magnetic field (*M-H*) curves measured at 10K, for spin-injectors with different CoFeB thicknesses annealed at $T_{an}$=250°C. From 1 nm to 1.2 nm, the saturation field quickly decreases and the remanence increases with the CoFeB thickness. However, when the CoFeB thickness is larger than 1.2 nm, the saturation field increases and the remanence decreases again. This is due to the fact that the CoFeB layer is discontinuous and exhibits a superparamagnetic character for thicknesses below 1.2 nm. Each CoFeB island has its own magnetization, and the magnetization direction can be randomly flipped under the influence of temperature. In the absence of an external magnetic field, the average magnetization is zero and no remnant magnetization can be measured. When the thickness exceeds 1.2 nm, CoFeB forms a continuous film exhibiting ferromagnetic properties. The contribution of the MgO/CoFeB interface anisotropy ($K_i$) can overcome the CoFeB bulk in-plane shape anisotropy ($K_b$) and establish the perpendicular magnetic anisotropy of the CoFeB layer[35]. Since the contribution of $K_i$ is inversely proportional to the CoFeB thickness, the magnetization of CoFeB turns back to in-plane for a large thickness of CoFeB.

To extract the interface anisotropy $K_i$, we need to study the effective anisotropy energy as a function of the CoFeB thickness. Since it is reported that there exists a magnetic dead layer which is attributed to the intermixing at the top Ta/CoFeB interface during deposition or upon annealing[43], we firstly extracted the values of the dead layer thickness ($t_d$) and of the saturation magnetization ($M_s$) from the fit of the magnetization as a function of the CoFeB thickness. As shown in **Figure 3b**, the $M_s$ value is extracted to be 1216 emu/cm$^3$. $t_d$ is estimated to be 0.05nm, which is much smaller than the reported value of 0.5nm[35], probably due to the specific deposition condition that we used. The effective anisotropy energy density per unit volume ($K_{eff}$) can be extracted from the integrated difference between the out-of-plane and in-plane *M-H* curves[44] and plotted *vs.* the effective CoFeB thickness $t_{eff}=t_{CoFeB}-t_d$, as shown in **Figure 3c**. Please see **Supplementary Note1 and Note2** for the details. The easy magnetization axis of the CoFeB layer is out-of-plane when $K_{eff}>0$ and in-plane when $K_{eff}<0$. The interface anisotropy ($K_i$) can be obtained from the intercept of $K_{eff}\cdot t_{eff}$ *vs.* $t_{eff}$ linear fitting. We find a value of 0.97±0.10 mJ/m$^2$, which is comparable to the value of 1.3 mJ/m$^2$ given by Ikeda *et al.* for a MTJ[31], and larger than the value of 0.63 mJ/m$^2$ reported for GaAs spin LEDs[35].

A precise control over the annealing temperature $T_{an}$ is also important to obtain a stronger PMA. **Figure 3d** displays the comparison of the out-of-plane *M-H* curves for the 1.2 nm thick CoFeB sample annealed at different temperatures. The corresponding $K_{eff}$ *vs.* $T_{an}$ is plotted in **Figure 3e**. Please see details in **Supplementary Note3**. The optimized annealing temperature is found to be around 300°C. The CoFeB coercivity and remanence gradually increase with the annealing temperature up to 300°C, then decrease above the annealing temperature of 350°C. As already investigated theoretically by Yang *et al.*[32], the PMA is very sensitive to the chemical structure of the Fe(Co)/MgO interface. The improvement of the PMA up to 300ºC could be attributed to an optimization of the interfacial chemical structure and to the starting crystallization of the CoFeB at the vicinity of the CoFeB/MgO interface[39,45]. When $T_{an}$ exceeds 300ºC, Ta atoms start to diffuse through the ultrathin CoFeB layer to the MgO interface and significantly damage the PMA[36]. **Figures 3f** and **3g** show the in-plane and out-of-plane *M-H* curves recorded at 10K and 300K respectively, for the sample with the optimized conditions of $t_{CoFeB}$=1.2 nm and $T_{an}$=300°C. We can observe that although the coercivity is much reduced at RT, the perpendicular magnetic anisotropy still persists up to RT.

To clarify the role of both $MoS_2$ and Ta layers in establishing the PMA, we have prepared two additional reference samples. One contains the same Ta/CoFeB/MgO structure grown directly on a $SiO_2$ substrate (*i.e.* without any $MoS_2$ layer) and the other one is grown on a $MoS_2$ ML, but with the Ta layer replaced by a Pt layer. **Figure 3h** displays the out-of-plane M-H curves recorded at 10K for the two samples after annealing at 300°C in comparison with the Ta/CoFeB/MgO/$MoS_2$ sample. It is found that the sample grown on $SiO_2$ shows a smaller remanence (0.5) and coercivity (20 mT), indicating a much reduced PMA in the CoFeB layer. Surprisingly, the sample covered with Pt displays a large saturation field (up to 0.7T) and almost zero remanence, which indicates that the CoFeB magnetization in this case is completely in-plane. Compared to the amorphous $SiO_2$ substrate, the crystalline $MoS_2$ ML favors the crystallization of a textured MgO (001) film, which can serve as a good template for the crystallization of CoFeB layer in *bcc* structure during the annealing procedure. This is an important prerequisite condition to create the large MgO/CoFeB interface anisotropy responsible for the PMA of CoFeB. Therefore, it is not surprising that the PMA in the reference sample without $MoS_2$ is much reduced, thus validating the important role of the $MoS_2$ ML.

However, it is difficult to understand the behavior of the sample capped with the Pt layer, because a PMA has already been reported for the Pt/CoFeB/$MoSe_2$ system[25]. To understand why the PMA is completely lost, we have also performed STEM-EELS analyses on the Pt capped sample. **Figure 4a** and **4b** display the large scale and magnified HRTEM images of the Pt/CoFeB/MgO/$MoS_2$ interfaces after annealing at 300°C, respectively. It is found that the MgO layer contains many small crystalline structures with disordered orientations. The CoFeB layer also exhibits crystalline features. The insets of **Figure 4b** display the FFT patterns recorded on three close zones of the CoFeB (marked with red dashed squares). The different diffraction patterns indicate that CoFeB layer is constituted of polycrystalline nanocrystals without any textured structures. We have also performed STEM-EELS analyses of element map on this sample. **Figure 4c** show the STEM HAADF image of the zone where EELS analyses are performed. **Figures 4d-i** show the chemical maps drawn after EELS spectrum image processing. The Mo signal is not shown due to some artificial effects in the Pt area. Compared to the sample with Ta capping layer, the most significant difference is the location of B revealed by the mapping of B element (**Figure 4g**). Indeed, it clearly appears that most of B atoms diffuse inside the MgO instead of being absorbed by the Pt layer after annealing. Moreover, an important diffusion of Fe atoms into the Pt layer can be also evidenced. More details concerning the profiles of the element distribution and the EELS spectrum of $B_K$ edge structure can be found in **Supplementary Note4**. Compared to Ta, the reaction of boron with Pt is more unlikely to occur because it requires high temperature and high pressure conditions[46], which impedes the absorption of B during the annealing procedure at 300°C. On the contrary, it is well known that Pt and Fe can easily react with each other to form an alloy[47], which can explain the observed important diffusion of Fe in Pt. Since the Pt layer is already textured in the as-grown state, it can serve as a template for the CoFeB crystallization during the annealing, leading to the misorientated (non *bcc*) structure of the CoFeB layer. This phenomenon has already been observed when a Ru layer is contacted with CoFeB[48]. Due to the crystallization of CoFeB starting from the interface CoFeB/Pt, B is rejected towards the MgO layer, and the B diffusion is favored by its oxidation inside MgO[49,50,51] (see in **Supplementary Note4**). The combined effects of the B diffusion inside MgO and the misorientated CoFeB crystallization completely destroy the PMA at the CoFeB/MgO interface, which in-turn explains the observed in-plane magnetization in the Pt capped sample. This study thus highlight the critical role of Ta to efficiently absorb B atoms from the CoFeB layer to establish the PMA.

*First principle calculations of the electronic structure of the Fe/MgO/MoS$_2$ multilayer*

To know the potentiality of such FM/oxide/2D hetero-junction for electrical spin injection, we have performed *ab-initio* calculations to investigate the band structure and spin resolved density of states (DOS) at the interfaces of a Fe/MgO/MoS$_2$ slab, which simulates our experimental CoFeB/MgO/MoS$_2$ structure. This simulation can allow us to get the information of the band alignment, spin-resolved DOS decay in the oxide barrier and band structure modification in the 2D material due to the FM layer, which are important for us to design our system for efficient spin injection.

**Figures 5a** shows the section and top view of the unit cell that we have used to model the Fe/MgO/MoS$_2$ multilayer. It contains 7 MLs of Fe, followed by 7 MLs of MgO and a single layer of 1H-MoS$_2$. The $O_z$ axis is the direction perpendicular to the Fe/MgO and MgO/MoS$_2$ interfaces. The conventional MgO cubic cell is rotated in the *xy*-plane by 45° with respect to the conventional Fe cubic cell, Fe atoms being located on top of oxygen atoms at the Fe/MgO interface[32]. The Fe/MgO/MoS$_2$ supercell that we used for the calculations imposes the same periodicity in *x*- and *y*-directions for the Fe, MgO and MoS$_2$ layers. This common periodicity may be responsible for structural distortions in the heterostructure unless huge supercells are used. Therefore, we had to find a compromise between reasonable distortions and reasonable supercell size. Please see **Methods** for more details about the construction of the slab. After relaxation of the atomic positions, a distance of 0.441 nm is found between the Mo layer and the first MgO layer (0.285 nm between MgO and the interface S layer, and ~0.156 nm between the Mo layer and each of the two S layers), confirming the weak Van-der-Waals bonding between MoS$_2$ and the rest of the stacking. This is also in a rather good agreement with the value of 0.4 nm measured from TEM observations.

**Figure 5b** shows the DOS integrated over atomic spheres belonging to the successive atomic layers of the supercell. The DOS curves calculated at the center of the Fe layer (Fe4) and at the center of the MgO layer (MgO4) resemble those calculated for undistorted *bcc* Fe and *fcc* MgO, respectively. This indicates that the artificial distortion that we imposed does not strongly change the physical properties of the Fe and MgO layers. The spin polarization at the Fermi level $E_F$, $P_s(E_F)$, is defined as the relative difference between the DOS values calculated for the two spin states at $E_F$. It is negative for the interface Fe ML (Fe7), the majority spin DOS being lower than the minority spin DOS. Most importantly, the bottom of the conduction band of MoS$_2$, which was clearly above the Fermi level $E_F$ for an isolated MoS$_2$ layer, is now shifted below $E_F$. It means that electrons have been transferred from the Fe/MgO interface to the MoS$_2$ layer and the resulting charge distribution induces an internal electric field in MgO (shown as a continuous shift of the successive MgO DOS curves). This internal electric field is partly screened by a small buckling (between 0.0025 and 0.005 nm) between Mg and O atoms of the same MgO layer. **Figure 5c,** which shows the *z*-variations of the *xy*-averaged electrostatic potential, confirms that the internal electric field only exists in the MgO layer. **Figure 5d** shows the modification of the electron charge when the 7 ML-thick Fe layer, 7 ML-thick MgO layer and MoS$_2$ monolayer are associated in the same supercell. The comparison with a similar figure calculated for the Fe(7 MLs)/MgO(7 MLs) bilayer (not shown) reveals that the charge transfer mainly occurs between the Fe layer at the Fe/MgO interface and the S layer at the MgO/MoS$_2$ interface. The charge transfer in this metal-insulator-semiconductor (MIS) structure is due to the different work functions of electrons in bulk Fe and in the MoS$_2$ ML. Our calculated values of the Fe work function and the MoS$_2$ ML electron affinity and band gap energy are indeed of 3.91 eV, 3.65 eV and 1.74 eV, respectively (while the values of 4.5 eV[52], 4 eV[53,54] and 1.8 eV[1] have respectively been measured in experiments). A similar charge transfer has also been experimentally observed in the Fe/MgO/GaAs system[55]. The charge transfer induced by the insertion of MgO can effectively reduce the Schottky barrier height for FM/2D contact[31,56,57], which is in favor of the spin-polarized electron injection from the FM metal to the 2D conduction band.

**Figure 5e** shows the *z*-dependence of the DOS at the Fermi level, calculated for majority and minority spin electrons. The DOS values decrease exponentially in MgO, both from the Fe/MgO and MgO/MoS$_2$ interfaces. **Figure 5f** shows the corresponding profile of spin polarization at $E_F$ calculated from the spin-resolved DOS. The spin polarization at the Fermi level rapidly vanishes away from the MgO/Fe interface, and the electron gas transferred in the MoS$_2$ layer does not show significant spin polarization. The sign of spin polarization is changed in the middle of the MgO layer. This is because the characteristic length of the exponential decay of electron states near the Fe/MgO interface is smaller for minority spin than for majority spin electrons, in agreement with the results obtained by Butler *et al*. for the Fe/non-distorted MgO interface[58].

To clarify the influence of the Fe/MgO stack on the band structure of $MoS_2$, we have compared the band structures of the $MoS_2$ ML and the Fe/MgO/$MoS_2$ multilayer. **Figure 6a** shows the rectangular first Brillouin zone of the Fe/MgO/$MoS_2$ multilayer (black lines) and the hexagonal Brillouin zone of the $MoS_2$ ML (red lines). To understand where the *K* and *K'* valleys of $MoS_2$ are located in the rectangular Brillouin zone, the pieces of the $MoS_2$ ML hexagonal Brillouin zone located outside the rectangular one must be translated inside the black rectangular zone: the location of the *K*, *K'* and *M* points after these translations are indicated with black letters. This helps to understand how the band structure of $MoS_2$ is folded, when represented along the high symmetry directions of the multilayer Brillouin zone. Please see **Supplementary Note5** where the band structure of the $MoS_2$ ML is represented in the hexagonal Brillouin zone, then folded in the rectangular one. **Figure 6b** shows the calculated band structure of the $MoS_2$ ML with a rectangular conventional cell, which clearly indicates a direct band gap of 1.7 eV at K and K' points. **Figure 6c** shows the contribution of the $MoS_2$ layer to the band structure of the multilayer containing 7 MLs of MgO. The band structure of $MoS_2$ confirms that electrons transferred from the Fe/MgO interface to the $MoS_2$ layer occupy the minima of the conduction bands of $MoS_2$ at the K and K' points. Furthermore, we find that the band gap of the $MoS_2$ ML is no longer direct. The maximum of the valence band is found at the Γ point instead of the K and K' points. The colors in this figure show the value of the spin-component $s_z$ for each of the Bloch states forming these bands. The value of $s_z$ vanishes near Γ for the valence band (where the contrast is lost), which means that the spin is oriented in the $MoS_2$ plane for these states. To better understand the origin of the modification of the band structure, we have also performed calculations in a similar structure containing only 3 MLs of MgO. Please see **Supplementary Note6** for more details. As shown in **Figure 6d**, we observe that $MoS_2$ still keeps its direct bandgap in the multilayer with 3 MLs of MgO (the top of the valence band at Γ being shifted to a lower energy), which shows that the influence of the MgO layer on the band structure of MoS$_2$ is not trivial and strongly depends on its thickness.

Another interesting feature is that a Zeeman splitting of about 10meV is evidenced at the Γ point for the Fe/MgO(3 MLs)/MoS$_2$ system (inset of **Figure 6d**), while this splitting is negligible for the MoS$_2$ ML (inset of **Figure 6b**) and the Fe/MgO(7 MLs)/MoS$_2$ system (inset of **Figure 6c**). The Zeeman splitting is attributed to a proximity effect due to the presence of ferromagnetic Fe on the other side of the MgO layer. The proximity effects have been also recently evidenced in the cases of CoFeB/MgO/Si junction by spin-pumping measurements [59] and CrI$_3$/WSe$_2$ systems by photoluminescence measurements[60]. It is quite interesting that the thickness of the inserted MgO layer plays a role on the magnetic properties of MoS$_2$ by the proximity effect. Compared to the case of the direct contact between FM metals and 2D materials, the insertion of a thin insulating layer can prevent the hybridization at the metal/TMD interface and enable room temperature controlled proximity effects[61]. For 2D spin optoelectronic applications, thick MgO barrier will enhance the spin filtering effect to increase the injected spin polarization[58]. However, the band gap of the MoS$_2$ layer will become indirect. On the contrary, with a thin MgO, MoS$_2$ preserves a direct band gap and the proximity effects will enhance the spin lifetime in MoS$_2$[62]. In the future, it will be very interesting to explore the interplay between proximity effect and spin injection efficiency with the perpendicularly magnetized spin injector in 2D spin optoelectronic devices.

**CONCLUSION**

We have reported a large perpendicular magnetic anisotropy in Ta/CoFeB/MgO structure on full coverage monolayer MoS$_2$. The large PMA has been obtained by a precise control of the CoFeB thickness and the post-annealing process. It is found that the insertion of a thin insulating MgO layer can effectively block the metal diffusion into the 2D material during the annealing treatment. MgO can be crystallized to the (001) *fcc* texture on the MoS$_2$ ML. During annealing at 300°C, the Ta capping layer shows a high efficiency to absorb B atoms from the CoFeB layer. This prevents the diffusion of B atoms into MgO and the suppression of the PMA, as evidenced by a similar sample capped with Pt. All these factors ensure the large PMA established in the Ta/CoFeB/MgO/MoS$_2$ system. Furthermore, first principle calculations have also been performed in similar Fe/MgO/MoS$_2$ structures to understand the electronic and spin properties in such FM/oxide/2D systems. It is of high interest to find that the MgO thickness can modify the MoS$_2$ band structure from an indirect bandgap with 7 MLs of MgO to a direct one with only 3 MLs of MgO. The latter multilayer also shows a proximity effect with a Zeeman splitting of 10 meV in the MoS$_2$ valence band at the Γ point. These findings suggest that the magnetic and electronic properties of MoS$_2$ can be modulated by adjusting the MgO thickness. Our experimental and theoretical results will promote the future development of room temperature spin optoelectronic devices based on 2D TMDs with a perpendicularly magnetized spin injector.

**Methods**

*Sample growth and preparation*

Monolayer MoS$_2$ was grown by CVD using highest purity (6N) gases and precursors on SiO$_2$ substrate (purchased from 2D semiconductor Inc.). The MoS$_2$ substrate is firstly introduced into a molecular beam epitaxy (MBE) system to perform an annealing at 150°C during one hour to desorb water and CO$_2$ on the surface. After cooling down to RT, the sample was transferred to another sputtering system (with a base pressure of 5×10$^{-8}$ torr) without breaking the vacuum. A multilayer stack consisting of MgO (1.5nm) /Co$_{0.4}$Fe$_{0.4}$B$_{0.2}$ (1.2nm) /Ta (5nm) was deposited at RT. After taking out the sample from the vacuum, we have performed rapid thermal annealing treatments at different temperatures for 3 min.

*STEM-EELS characterization*

HR-STEM combined with spatially resolved EELS was performed by using a probe-corrected microscope JEOL ARM200F (cold FEG) equipped with a GATAN GIF quantum energy filter to reveal the structure and element distribution in the spin-injector after annealing. The microscope was operated at 80 kV. High angle annular dark-field (HAADF), annular dark-field (ADF) and bright-field (BF) images were simultaneously recorded for investigating the heterostructure while only HAADF signal was recorded during EELS mapping. EELS spectrum images (SI) were recorded with a probe current of about 50 pA. Two EELS-SI were simultaneously recorded: one for the low-loss part containing the zero-loss, the other for the core loss, which allows advanced data post processing (correction of energy drift, multiple scattering corrections). A multivariate statistical analysis software (temDM MSA) was used to improve the quality of the STEM-EELS data by de-noising the core-loss SI before its processing to draw quantitative chemical maps [63]. Thin lamellas were extracted by focused ion beam (FIB) milling using an FEI Helios Nanolab 600i dual beam.

*First principle calculation*

The electronic structure of the multilayers has been calculated with the Vienna *ab initio* simulation package (VASP)[64,65], based on the density-functional theory (DFT) and the projector augmented-wave (PAW) method[66]. The generalized-gradient approximation developed by Perdew, Burke and Ernzerhof (PBE) was used to calculate the exchange and correlation potential[67]. We chose a basis size corresponding to a cut-off energy of 550 eV (which allows to calculate the ground state energy with a precision of 0.0025 eV for the $MoS_2$ monolayer, 0.05 eV for the Fe slab and 0.05 eV for the MgO slab). The first Brillouin zone was sampled with a 20×16×1 Monkhorst-Pack grid[68] for structural optimizations and for calculating the band structure. This corresponds to a k-point spacing of 1 $nm^{-1}$ along x and 0.7 $nm^{-1}$ along y, which ensures that the ground state energy is calculated with a precision of 0.015% (0.044 eV). A denser mesh (45×45×1) is used to calculate the density of states (k-point spacing of 0.44 $nm^{-1}$ and 0.28 $nm^{-1}$ along x and y, respectively). The atom internal coordinates have been optimized until all forces are below 10 meV/Å and Van-der-Waals corrections have been taken into account within the DFTD3 formalism[69,70] (which gives values for the Van-Der-Waals gap of bulk $MoS_2$ in very good agreement with experiments). A dipole correction has been used to avoid non-physical charge transfer across the slab. We used the PYPROCAR library for the electronic structure pre/post-processing, in particular to plot the band structures. The spin-orbit coupling has been taken into account for this calculation.

Since the characteristic dimensions of the cubic Fe and MgO lattices ($a_{Fe} = 0.287$ nm and $a_{MgO} = 0.421$ nm) do not naturally match those of the hexagonal MoS₂ one ($a_{MoS_2} = 0.317$ nm, according to experiments), we chose to preserve the atomic structure of the MoS₂ layer and to artificially slightly distort that of the Fe and MgO layers, until the matching of the atomic structures is reached. This occurs when the distortion transforms the MgO squares of the (001) MgO layers with dimensions $\frac{a_{MgO}}{\sqrt{2}} \times \frac{a_{MgO}}{\sqrt{2}}$ into rectangles with dimensions $a_{MoS_2} \times \frac{\sqrt{3}}{2} a_{MoS_2}$. Compatible distortions have been applied to the Fe layers. For MgO, this corresponds to strains of 6.5% in the x-direction and -7.8% in the y-direction. For Fe, the strains are of 10.4% and -4.3% along x and y. The size of the unit cells was fixed to $a_{MoS_2}$ and $\sqrt{3} a_{MoS_2}$ in the *x*- and *y*-directions, where $a_{MoS_2} = 0.316$ nm is the value calculated from first principles. The unit-cell length in the *z*-direction was adjusted such that the surface Fe layer and the periodic image MoS₂ layer are separated by 1.6 nm of vacuum.

**Supporting Information.** Procedure to extract effective anisotropy energy Keff; Details for extracting Keff for the samples with different thickness of CoFeB; Details for extracting Keff for the samples with different annealing temperature; Element distribution profile and B diffusion in MgO for the sample capped with Pt; Band structure of the primitive (hexagonal) and conventional (rectangular) unit cells of a MoS2 ML; First principle calculations for the MoS2/3 MLs MgO/7 MLs Fe structure.

**Acknowledgement**

We acknowledge Stéphane Suire for the support of magnetism measurement from the magnetism center of Institut Jean Lamour. This work is supported by ANR SIZMO2D project (Grant No. ANR-19-CE24-0005) and ANR FEOrgSpin project (Grant No. ANR-18-CE24-0017). We also acknowledge ICEEL (international) SHATIPN project and CPER MatDS project. Experiments were performed using equipments from the platform TUBE-Davm and the platform CCMEM both funded by FEDER (EU), ANR, the Region Lorraine and Grand Nancy. Z.W. acknowledges the support of the National Key Research and Development Program of China (Grant No. 2017YFA0207500). This work was granted access to the HPC resources of CALMIP supercomputing center under the allocation 2020/2021-[P20042].

**Figures**

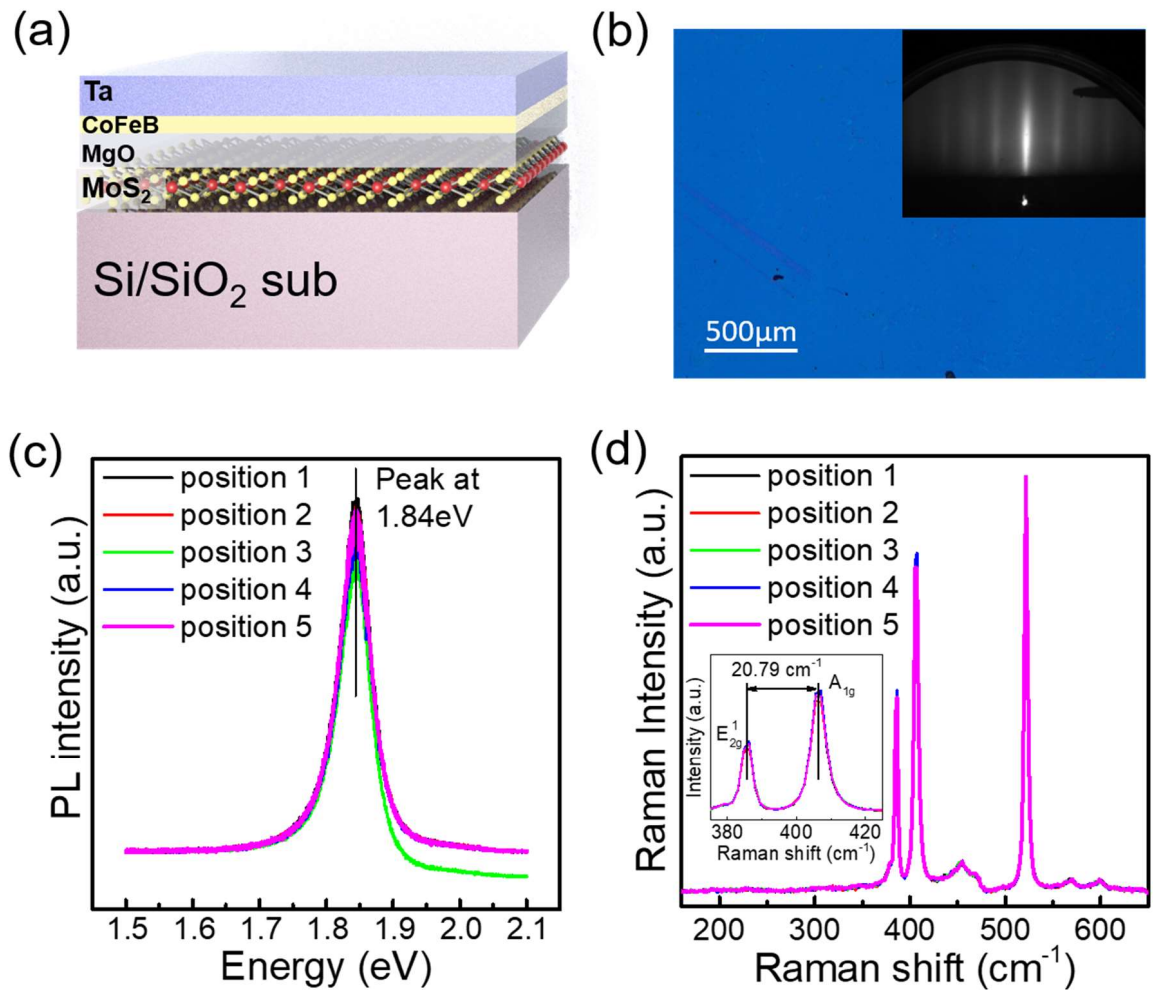

Figure 1: Schematic structure of the investigated sample and characterization of the MoS$_2$ ML. (a) Schematics of the Ta/CoFeB/MgO/MoS$_2$ stack on Si/SiO$_2$ substrate. (b) Large scale optical microscopy image of the surface of monolayer MoS$_2$. Inset: RHEED pattern of the MoS$_2$ surface after annealing at 150°C for one hour. The pattern does not change with different angles of incident e-beam. (c) Room temperature photoluminescence spectra taken on different positions of the MoS$_2$ ML surface. (d) Raman spectra taken on different positions of the MoS$_2$ ML surface. Inset: Enlarged spectra showing the distance between $E_{2g}^1$ and $A_{1g}$ modes.

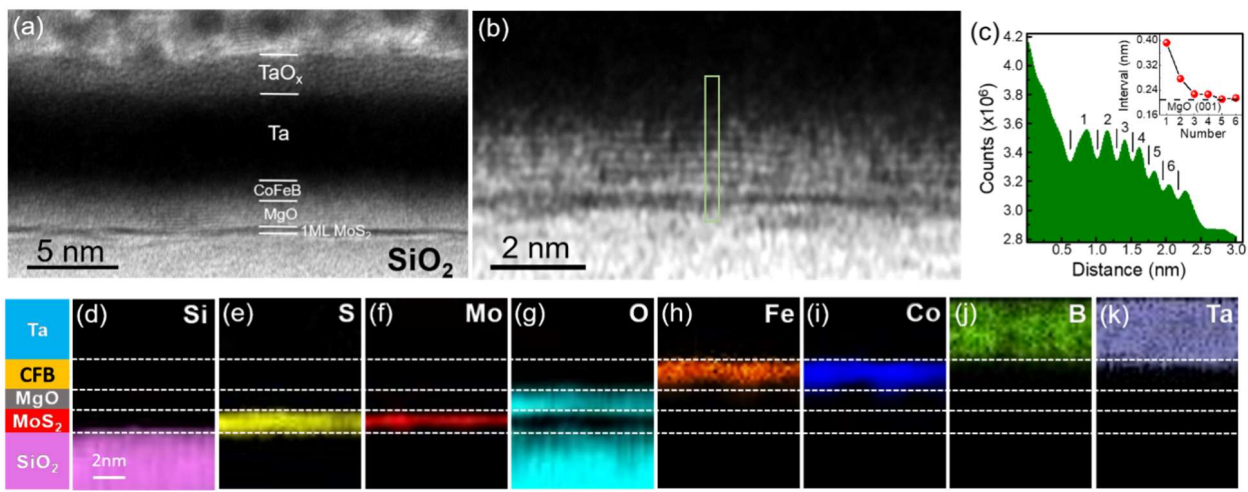

Figure 2: Interfacial structure and chemical characterization of the Ta/CoFeB/MgO/MoS$_2$ heterostructure. (a) Large scale HR-STEM bright field image showing the multilayer structure. (b) Enlarged image showing the crystallization of MgO on MoS$_2$. (c) Integrated intensity profile along the green area in panel (b). Inset: Distance separating two successive oscillation minima as a function of the interval number. (d-k) Chemical maps drawn by processing EELS spectrum images using the signals of (d) Si$_{L3}$ (99 eV), (e) S$_{L3}$ (165 eV), (f) Mo$_{M5}$ (227 eV), (g) O$_K$ (532 eV), (h) Fe$_{L3}$ (708 eV), (i) Co$_{L3}$ (779 eV), (j) B$_K$ (188 eV) and (k) Ta$_{N5}$ (229 eV) edges, respectively.

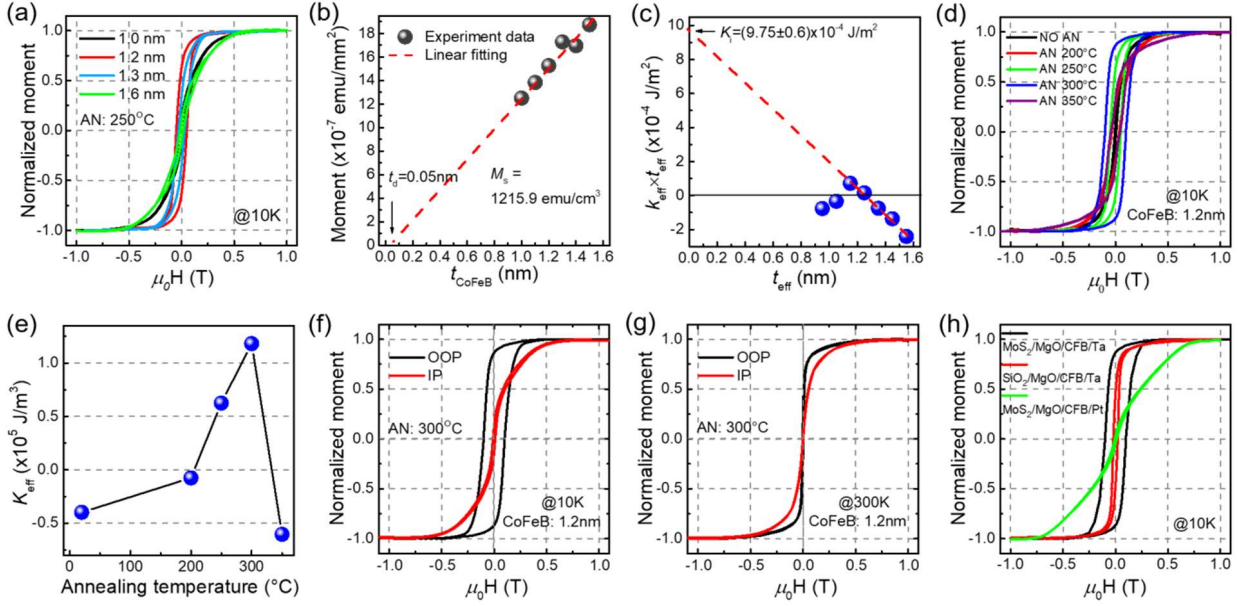

Figure 3: Magnetic characterization of Ta/CoFeB/MgO/MoS$_2$ heterostructures. (a) Out-of-plane *M-H* curves for samples with different CoFeB thicknesses with $T_{an}$=250°C measured at 10K. (b) Extrapolation of CoFeB magnetic dead layer $t_d$ from the CoFeB thickness dependent saturation magnetization $M_s$. (c) $t_{eff}$ dependence of the product of $K_{eff}$ and $t_{eff}$, where the intercept to the vertical axis of the linear extrapolation corresponds to the interface anisotropy $K_i$. (d) Out-of-plane *M-H* curves for the sample with 1.2nm CoFeB before and after different temperature annealing measured at 10K. (e) $K_{eff}$ as a function of annealing temperature. (f,g) *M-H* curves for the sample with 1.2nm CoFeB and $T_{an}$=250°C for in-plane and out-of-plane configurations measured at (f) 10K and (g) 300K. (h) Comparisons of out-of-plane *M-H* curves for three samples: Ta/CoFeB/MgO/MoS$_2$, Ta/CoFeB/MgO/SiO$_2$ and Pt/CoFeB/MgO/MoS$_2$. All samples have the same CoFeB thickness and were annealed at 300°C.

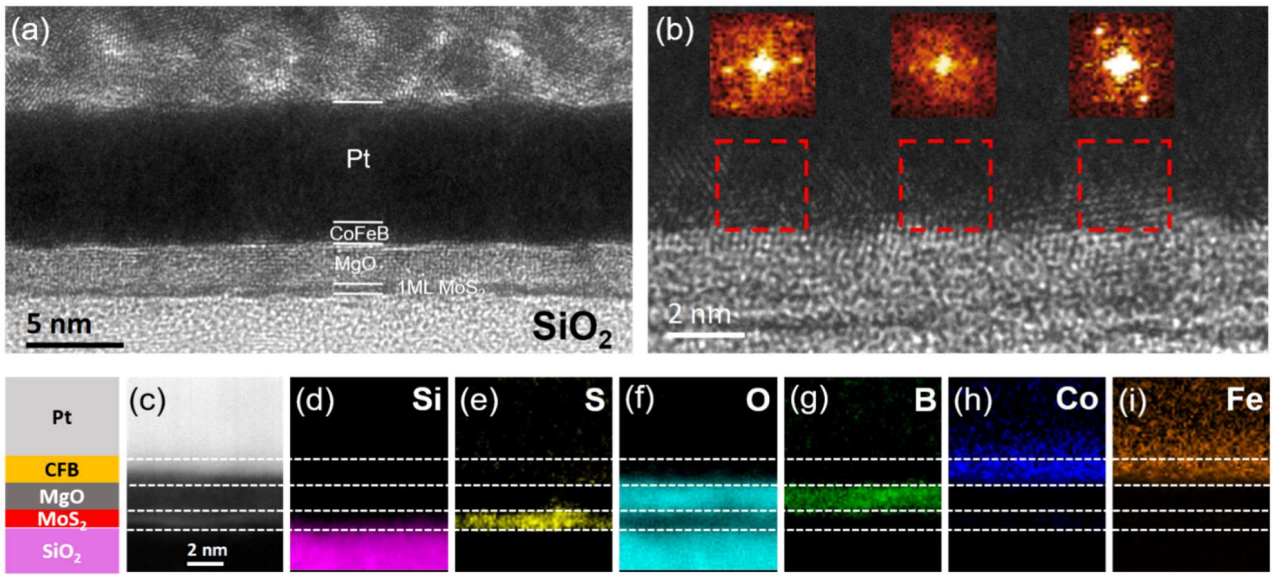

Figure 4: Interfacial structure and chemical characterization of Pt/CoFeB/MgO/MoS$_2$ heterostructure. (a) Large scale HR-TEM image showing the multilayer structure. (b) Enlarged image showing the crystallization of CoFeB on MgO. Insets: FFT patterns showing that the CoFeB is characterized by polycrystalline features (here 3 different orientations for very close zones). (c) STEM HAADF image of the zone where EELS analyses are performed. (d-i) Element maps drawn from STEM-EELS spectrum image, respectively for (d) Si$_K$, (e) S$_L$, (f) O$_K$, (g) B$_K$, (h) Co$_L$ and (i) Fe$_L$ edges.

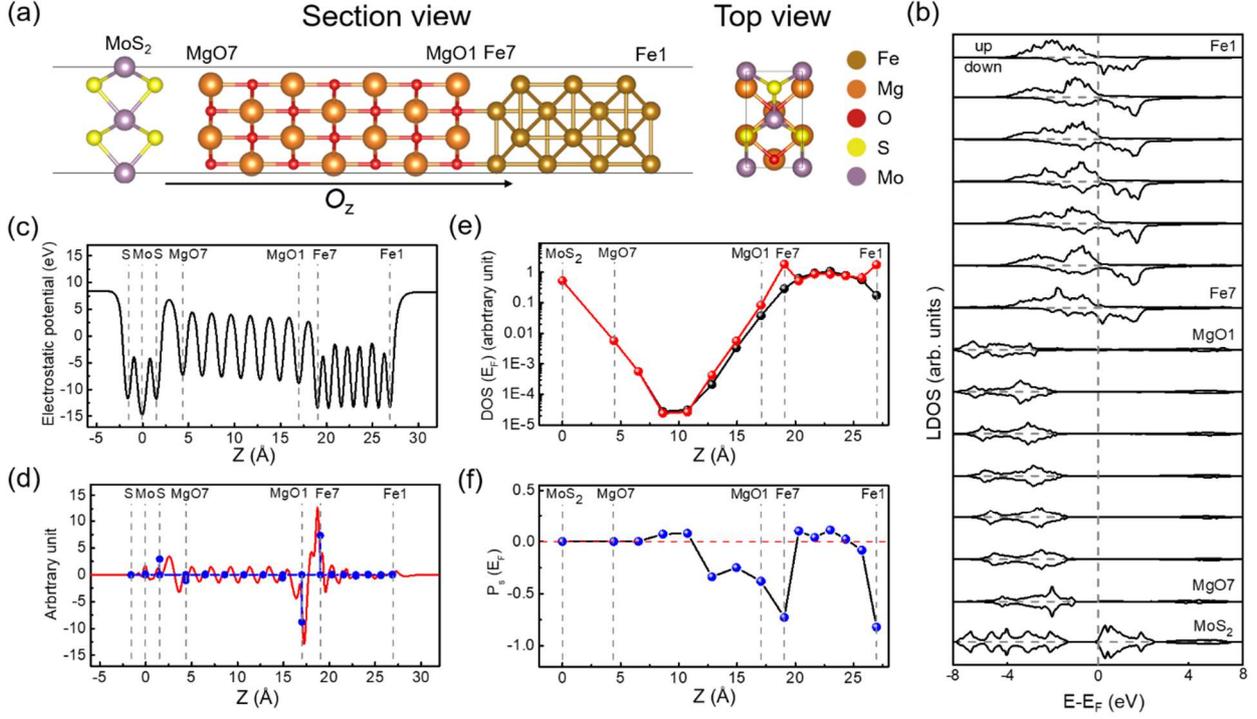

Figure 5: First principle calculations of the spin-resolved density of states. (a) Section and top view of the unit cell used to model the Fe/MgO/MoS$_2$ multilayers. (b) Majority and minority spin DOS integrated over atomic spheres belonging to the successive atomic layers of the supercell. (c) $z$-variations of the $xy$-averaged electrostatic potential. (d) The red curve corresponds to the variations of the electron charge density when the 7 ML-thick Fe, 7 ML-thick MgO and MoS$_2$ ML are associated in the same supercell. Each blue dot corresponds to the $z$-integration of the red curve between the two mid-perpendicular planes separating an atomic layer and its neighbors on the right and left sides. (e) $z$-dependence of the DOS at the Fermi level, calculated for majority spin (black curve) and minority spin (red curve) electrons. (f) Profile of spin polarization at the Fermi level, calculated from the spin-resolved DOS.

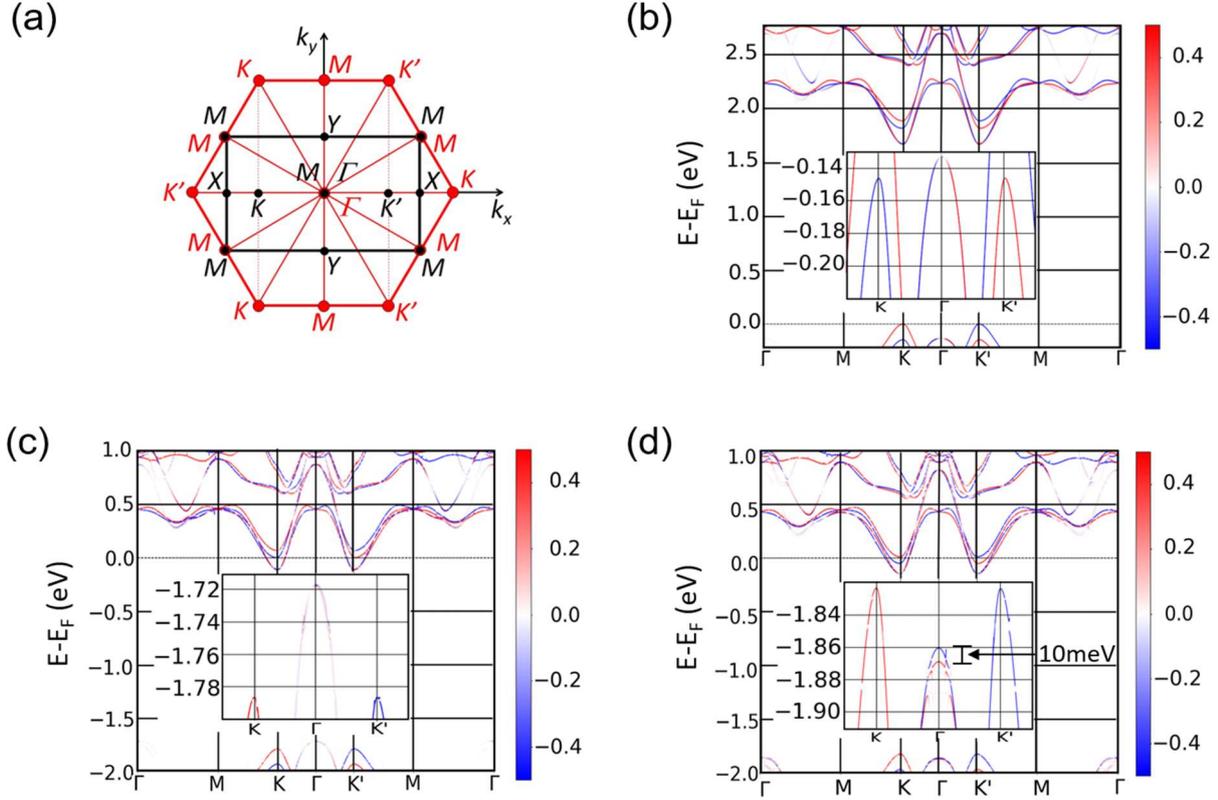

Figure 6: Band structure of the MoS$_2$ ML in the multilayer structure. (a) Brillouin zone of the unit cells. The red lines represent the hexagonal Brillouin zone of the primitive unit cell of the MoS$_2$ ML. The black lines represent the rectangular Brillouin zone of the multilayer structure. (b) Band structure of monolayer MoS$_2$ calculated with a conventional rectangular unit cell. Inset: Magnified valence band structures. The color bar shows the value of the spin-component $s_z$ for each of the Bloch states forming these bands. (c,d) Contribution of the MoS$_2$ layer to the band structure of the multilayer with (c) 7 MLs of MgO and (d) 3 MLs of MgO. The red and blue lines respectively show the positive and negative values of the spin projection $s_z$ (in units of $\hbar$). Insets: Magnified valence band structures show a neglectable Zeeman splitting for 7 MLs of MgO but a large splitting of 10mV at Γ point for 3 MLs of MgO.

**TOC figure**

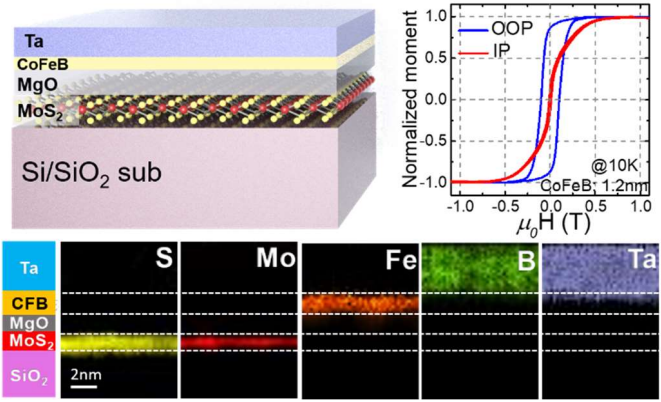